\documentstyle[preprint,aps]{revtex}
\begin{document}
\draft
\preprint{}
\title{Ultra-Relativistic Hamiltonian with Various Singular Potentials}
\author{Atsushi Yoshida}
\address{Department of Physics, University of Virginia, 
Charlottesville, VA 22901}
\date{\today}
\maketitle
\begin{abstract}
It is shown from a simple scaling invariance that the ultra-relativistic
Hamiltonian ($\mu$=0) does not have bound states when the potential is 
Coulombic.  This supplements the application of the relativistic
virial theorem derived by Lucha and Sch\"{o}berl \cite{{LuSch1},{LuSch2}} 
which
shows that bound states do not exist for potentials more singular than
the Coulomb potential.
\end{abstract} 

\newpage
\narrowtext

The relativistic generalization of the Schr\"{o}dinger equation (RSE)
\begin{equation}
\label{rse}
\sqrt{{\bf p}^2 + {\mu}^2} \psi({\bf x}) + V(r) \psi({\bf x}) =
(E+{\mu}) \psi({\bf x})
\end{equation}
has been used in describing quark-antiquark bound states when one of 
the constituents is light and the other heavy \cite{DuEiTh}.  The mass
$\mu$ can be considered as the constituent mass of the light quark and
$E$ is the binding energy.

This Hamiltonian has two 
interesting critical behaviors concerning the existence of bound states
when the potential is Coulombic, 
\begin{equation}
V_{c} = - {\alpha}/r,
\begin{array}{c} $ \space \space $ \alpha > 0.
\end{array}
\end{equation}
First, it has
been shown that, irrespective of the value of $\mu$, when 
the coupling constant of the Coulomb potential
$\alpha$ approaches the value $2/{\pi}$ from below, the bound
states disappear due to the large singularity of the potential at the 
origin \cite{Herbst}.  This means, as a corollary, that there are 
no bound states when the potential is more singular than Coulomb. 
In this letter, I show the second critical behavior concerning the
existence of bound states for a Coulombic potential: irrespective of 
the value of
coupling constant, bound states do not exist for a massless particle
($\mu = 0$).  This second critical behavior in terms of $\mu$ implies
(similar to the what is implied by the first critical behavior in
terms of $\alpha$) that there are no bound states with massless
particles for a potential more singular than Coulomb.

First I derive a scaling feature of RSE
for a general potential. Then I
concentrate on the Coulomb potential and show that bound states do not
exist for massless particles when the potential is Coulombic.  
I start with two wavefunctions
${\psi}({\bf x})$ and ${\tilde \psi}({\bf x})$ and their Fourier
transforms ${\phi}({\bf p})$ and ${\tilde \phi}({\bf p})$ related in 
the
following way,
\begin{equation}
\label{wf}
{\psi} ({\bf x}) \equiv {\tilde \psi} (t{\bf x}) 
\end{equation}
\begin{eqnarray}
{\phi} ({\bf p}) &=& {1 \over ({\sqrt {2\pi}})^3} 
\int_{- \infty}^{\infty} {\psi} ({\bf x}) 
e^{i{\bf p} \cdot {\bf x}} dx^3  \nonumber\\
&=& {1 \over t^3} {1 \over ({\sqrt {2\pi}})^3}
\int_{- \infty}^{\infty} {\tilde \psi} (t{\bf x}) e^{i {{\bf p} 
\over t} \cdot (t{\bf x})} d(tx)^3 \nonumber\\
&=& {1 \over t^3} {\tilde \phi} ({{\bf p} \over t}).
\end{eqnarray}
Here $t$ can be any positive real number.
Now suppose ${\psi}({\bf x})$ satisfies RSE, then
\begin{eqnarray}
\label{tilde}
(E+{\mu}) {\psi}({\bf x}) &=& 
\sqrt {{\bf p}^2 + {\mu}^2} {\psi} ({\bf x}) + V({\bf x}) 
{\psi} ({\bf x})
\nonumber\\
&=& {1 \over ({\sqrt {2\pi}})^3}
\int dp^3 \sqrt{{\bf p}^2 + {\mu}^2} {\phi} ({\bf p}) 
e^{-i{\bf p} \cdot {\bf x}} + V({\bf x}) {\psi} ({\bf x}) 
\nonumber\\
&=&{1 \over t^3} t^4 {1 \over ({\sqrt {2\pi}})^3}
\int d \Big{(} {p \over t} \Big{)}^3 \sqrt { \Big{(} 
{{\bf p} \over t} \Big{)}^2 + \Big{(} {{\mu} \over t} \Big{)}^2 } 
{\tilde \phi}
({{\bf p} \over t}) e^{-i{{\bf p} \over t} \cdot (t{\bf x})} + V({\bf x}) 
{\tilde \psi} (t{\bf x}) \nonumber\\
&=& t \sqrt {{\bf p}^2 + \Big{(} {{\mu} \over t} \Big{)}^2} 
{\tilde \psi} (t{\bf x}) 
+ V({\bf x}) {\tilde \psi} (t{\bf x})
\end{eqnarray}
Thus, using the equivalence (\ref{wf}), it gives
\begin{equation}
\label{general}
\sqrt {{\bf p}^2 + \Big{(} {{\mu} \over t} \Big{)}^2 } 
{\tilde \psi} ({\bf x}) 
+ {V({{\bf x} \over t})\over t} {\tilde \psi}({\bf x}) = \Big{(} 
{E \over t} 
+ {{\mu} \over t} \Big{)} {\tilde \psi} ({\bf x}).
\end{equation}
If the potential is Coulomb $V({\bf x}) = - {\alpha}/r$,  
then we arrive at the scaling results that if 
\begin{equation}
\label{eq1}
\sqrt {{\bf p}^2 + {\mu}^2} {\psi} ({\bf x}) - {\alpha
\over r}
{\psi} ({\bf x}) = (E+{\mu}) {\psi}({\bf x}) 
\end{equation}
then
\begin{equation}
\label{eq2}
\sqrt {{\bf p}^2 + {{\tilde \mu}}^2} {\tilde \psi ({\bf x})} 
- {\alpha \over r}
{\tilde \psi} ({\bf x}) = ({\tilde E}+{\tilde \mu}) {\tilde \psi }
({\bf x}),
\end{equation}
with the relations 
\begin{eqnarray}
\label{psi}
{\tilde \psi}({\bf x}) &=& {\psi} ({{\bf x} \over t}) \\
\label{mu}
{\tilde {\mu}} &=& {{\mu} \over t} \\
\label{e}
{\tilde E} &=& {E \over t}.
\end{eqnarray}

In fact this theorem applies to slightly more general potentials.  As
long as the parameters of the potential $V({\bf x})$ are all
dimensionless, then
\begin{equation}
\label{potcon}
{1 \over t} V({{\bf x} \over t}) = V({\bf x}).
\end{equation}
This is the only criteria that the potential must satisfy.  
In particular, the potential need not be spherically symmetric.
For example potentials like
\begin{equation}
\begin{array}{c}
V({\bf x}) = - {\alpha_1 \over \sqrt {x^2 + y^2}} - {\alpha_2 \over |z|}
$ \space \space \space \space (cylindrical) $
\end{array}
\nonumber
\end{equation}
and
\begin{equation}
\begin{array}{c}
V({\bf x}) = -{\alpha \over \sqrt{ax^2 + by^2 + cz^2}} $ \space \space
\space \space (ellipsoidal) $
\end{array}
\end{equation}
do satisfy eq.(\ref{potcon}).  The underlying reason for this generality is
that if the potential has only dimensionless parameters, the wavefunction is
restricted to the form
\begin{equation}
\label{f}
\psi({\bf x}) = {\mu}^{3 \over 2} f({\mu}{\bf x};\{ \alpha_i \}).
\end{equation}
Then after normalization, eq.(\ref{psi}) reduces to the statement
\begin{equation}
\begin{array}{c} $ if \space \space $ {\psi}({\bf x}) = {\mu}^{3 \over 2} 
f({\mu}{\bf x};\{ \alpha_i \}) $ \space \space \space 
 then  \space \space $ {\tilde \psi}({\bf x})
= {\tilde \mu}^{3 \over 2} f({\tilde \mu}{\bf x};\{ \alpha_i \}).
\end{array}
\end{equation}
Here the normalization is $\int dx^3 |f({\bf x};\{ \alpha_i \})|^2 
= 1$.  

In fact, from a dimensional point of view alone, one can show that
eq.(\ref{f}) must hold for non-relativistic Schr\"{o}dinger equation
(NRSE), and from this, all
the relations (\ref{psi}),(\ref{mu}) and (\ref{e}) as well.

This scaling behavior indicates, in words,
that as the mass increases by a factor of $t$, 
the wavefunction shrinks by a factor of
$t$ and the binding energy is also increased by a factor of $t$ 
(in magnitude).  From
this, it is clear that for ${\tilde \mu}=0$,  $\mu = 0$ automatically
as well.  This means now the Hamiltonians (\ref{eq1}) and (\ref{eq2})
become identical.  Since this is true for any arbitrary $t$, it follows
${\tilde E} \rightarrow -\infty$ as $t \rightarrow 0$.  Hence there 
are no massless bound states.  

One can also extract simple scaling
laws for other types of potentials from eq.(\ref{general}).  In 
particular,
for massless case with a potential $V(r) = -r^{-k}$ with $k>1$, one can
again show that bound states do not exist by taking the limit $t
\rightarrow 0$.  This time the potential is not invariant but
\begin{equation}
\label{v}
{\tilde V}({r \over t}) = t^{k} V(r).  
\end{equation}
Therefore, as $t \rightarrow 0$ the potential becomes shallower,
nevertheless, still the wavefunction shrinks to zero width and 
the energy goes to negative infinity.  Thus again there are no bound
states.  This is reminiscent to the well-known theorem in NRSE
which states that for a potential more singular
than $r^{-2}$ bound states do not exist \cite{Landau}.  Here in our 
situation, since 
massless bound states do not exist for a Coulomb potential
it is intuitively clear that for a potential more singular, the
same is true.

This fact that the bound states do not exist for a potential more
singular than the Coulomb potential can be inferred also from the
relativistic virial theorem (RVT) derived by Lucha and Sch\"{o}berl
\cite{{LuSch1},{LuSch2}}.  
It states, for an eigenstate of two-body relativistic
Hamiltonian (in the center-of-mass frame)
\begin{equation}
\label{ham}
H = \sqrt{{\bf p}^2 + m_{1}^{2}} + \sqrt{{\bf p}^2 + m_{2}^{2}} + V(x),
\end{equation}
the gradient of the potential is related to the kinetic energy as
\begin{equation}
\label{rvt1}
\langle {\bf x} \cdot \nabla V({\bf x}) \rangle 
= \langle {{\bf p}^2 \over \sqrt{{\bf p}^2 + m_{1}^{2}}} 
+ {{\bf p}^2 \over \sqrt{{\bf p}^2 + m_{2}^{2}}} \rangle.
\end{equation}
This leads to
\begin{equation}
\label{rvt2}
\varepsilon = \langle {\bf x} \cdot \nabla V({\bf x}) \rangle 
+ \langle V({\bf x}) \rangle 
+ \langle {m_{1}^2 \over \sqrt{{\bf p}^2 + m_{1}^{2}}} 
+ {m_{2}^2 \over \sqrt{{\bf p}^2 + m_{2}^{2}}} \rangle
\end{equation} 
where $\varepsilon$ denotes the total energy of the two-body system of
particle mass $m_1$ and $m_2$.  The Hamiltonian (\ref{ham})
simplifies to
(\ref{rse}) when $m_{2}$ is taken to infinity and $m_{1}$ is set to be
$\mu$.  
Accordingly (\ref{rvt2}) also reduces to
\begin{equation}
E = \langle {\bf x} \cdot \nabla V({\bf x}) \rangle 
+ \langle V({\bf x}) \rangle  
+ \langle {{\mu}^2 \over \sqrt{{\bf p}^2 + {\mu}^2}} \rangle. 
\end{equation}
Here the lhs has been reduced to the binding energy $E$.  
For a radially symmetric power law potential
\begin{equation}
V(r) = \alpha r^{k}
\end{equation}
where $\alpha$ is positive for $k > 0$ and negative for $k < 0$, 
this means simply
\begin{equation}
E = (k+1) \langle V \rangle 
+ \langle {{\mu}^2 \over \sqrt{{\bf p}^2 + {\mu}^2}} \rangle. 
\end{equation}
Clearly for $k < -1$, if bound states were to exist, it would give a
nonsensical result because the lhs is negative while rhs is positive.
This indicates that the eigenstates themselves do not exist for those
potentials with $k < -1$ for both finite and zero $\mu$.  
Literally taken, RVT predicts that all the massless particle bound
states have one and the same binding energy $E = 0$ for a Coulomb
potential ($k = -1$).  This could also
be seen as a manifestation that the bound states do not exist 
and the expectation values can not be defined.

Incidentally, in the 
non-relativistic case, not only these scaling relations
eqs.(\ref{psi}), (\ref{mu}), (\ref{e}) and (\ref{v}) hold, but more
general scaling relations can be derived.  Indeed, NRSE with a potential
of the form $V(r) = {\alpha}r^k$, where $\alpha$ is positive when $k>0$,
and negative when $k<0$, can be converted in radially reduced form to
the following dimensionless equation \cite{QuiRos},
\begin{equation}
-{d^2 \over d{\rho}^2}w({\rho}) + [sgn({\alpha}) {\rho}^k + {l(l+1)
\over {\rho}^2}] w({\rho}) = {\epsilon} w({\rho}).
\end{equation}
From this, it can be seen that 
all the scaling relations derived here are only 
a part of this more general scaling transformation.  
Unfortunately,
for the RSE case, this general transformation to the dimensionless form, 
which would allow one to extract a lot more informations on the
bound states, does not seem to be possible. 

I thank D.Singleton for reading the manuscript.

\end{document}